\documentclass[prl,reprint,twocolumn,showpacs,superscriptaddress]{revtex4-1}
\usepackage{amsmath,amssymb,bm,mathrsfs,graphicx, braket, times,color}
\usepackage[colorlinks=true,citecolor=blue,linkcolor=blue]{hyperref}
\begin{document}

\title{Energy gap of neutral excitations implies vanishing charge susceptibility}

\author{Haruki Watanabe}
\affiliation{Department of Applied Physics, University of Tokyo, Tokyo 113-8656, Japan.}

\begin{abstract}
In quantum many-body systems with a U(1) symmetry, such as the particle number conservation and the axial spin conservation, there are two distinct types of excitations: charge-neutral excitations and charged excitations. The energy gaps of these excitations may be independent with each other in strongly correlated systems.  The static susceptibility of the U(1) charge vanishes when the \emph{charged} excitations are all gapped, but its relation to the \emph{neutral} excitations is not obvious.  Here we show that a finite excitation gap of the \emph{neutral} excitations is, in fact, sufficient to prove that the charge susceptibility vanishes (i.e. the system is incompressible).  This result gives a partial explanation on why the celebrated quantization condition $n(S-m_z)\in\mathbb{Z}$ at magnetization plateaus works even in spatial dimensions greater than one.
\end{abstract}

\maketitle

\paragraph{Introduction.\,\,\,---}
When do we expect a plateau in a magnetization curve?  A very simple `quantization condition' is known to explain actual experiments over wide variety of materials in one~\cite{Narumi,Kikuchi,Zhangzhen,Hase}, two~\cite{Kageyama,Onizuka,Ono,Shirata,Ishikawa}, and three~\cite{Shiramura,Ueda,Ueda2} spatial dimensions: in spin models, a plateau appears when the $z$-component of the total spin is conserved and the magnetization per unit cell $m_z$ satisfies $S-m_z\in\mathbb{Z}$, where $S$ is the saturation magnetization per unit cell~\cite{OshikawaYamanakaAffleck,Fledderjohann,Oshikawa,Richter}. (For theoretical works on specific models, see references in Ref.~\cite{Richter}.)  When the unit cell is enlarged by spontaneous breaking of translation symmetry, $S$ and $m_z$ should be computed with respect to the `new' unit cell.  The reasoning leading to this condition is as follows.  It is natural to expect a finite excitation gap in the energy spectrum at a plateau.  In one-dimension, when $S-m_z$ is not an integer, one can construct a low-energy state by acting ``the twist operator" $\hat{U}=\exp\left[i\frac{2\pi}{L}\sum_{j=1}^Lj(S-\hat{S}_j^z)\right]$ on the ground state $|\Phi_0\rangle$~\cite{OshikawaYamanakaAffleck}, just as Lieb, Schultz, and Mattis did to constrain the low-energy spectrum of the antiferromagnetic chain~\cite{LiebSchultzMattis}. The energy expectation value of the state $\hat{U}|\Phi_0\rangle$ is bounded above by $O(L^{-1})$.  Therefore, in order to realize a gapped phase (without further enlarging the unit cell), one needs the above quantization condition. 

In higher dimensions, the system may avoid gapless excitations even when $S-m_z\not\in\mathbb{Z}$ without breaking translation symmetry by developing a `topological order.' As a consequence, a magnetization plateau may be formed at a `unquantized' $S-m_z$.  For example, a possible gapped spin liquid phase on the Kagome lattice~\cite{White,Depenbrock,Balents,Imai}, on the square lattice~\cite{Hong-Chen}, or more generally on a lattice that contains an odd number of spin-$1/2$s, if they are really gapped, will have a magnetization plateau at $m_z=0$ although $S-m_z$ is a half-odd integer.  Yet, the topological order is a single known exception to the quantization condition and otherwise one can discuss the necessity of $S-m_z\in\mathbb{Z}$ by extending the 1D argument~\cite{Oshikawa}.

There is, however, a loophole in the derivation of the quantization condition, as carefully remarked in the original works~\cite{OshikawaYamanakaAffleck, Oshikawa}.  As we will review shortly, those excitations relevant for the magnetization curve have different total magnetizations from the ground state.  The above low-energy state $\hat{U}|\Phi_0\rangle$, however, has the same total magnetization as $|\Phi_0\rangle$ does, because the twist operator $\hat{U}$ commutes with the $z$-component of the total spin $\sum_j\hat{S}_j^z$.  The quantization condition is thus only related to an excitation gap in the same magnetization sector as the ground state, but it does not tell anything about the excitations that change the total magnetization.  
Although there has been several improvements of the Lieb-Shultz-Mattis theorem in recent years~\cite{Affleck1986,Yamanaka,HastingsLSM,Sid,PNAS}, this point remains unchanged.
In one dimension, Ref.~\onlinecite{OshikawaYamanakaAffleck} developed an argument based on the Abelian bosonization to fill this gap to some extent, but the technique is essentially restricted to 1D.  Alternatively, when the system has the SU(2) spin rotation invariance, one can twist the ground state by $\hat{S}_j^x$ or $\hat{S}_j^y$, instead of $\hat{S}_j^z$, and the corresponding twist operator can produce a low-energy state with different magnetizations.
However, the spin rotation symmetry is usually broken down to the U(1) symmetry by the external magnetic field in the experimental setup.  Therefore, we have to explain why the quantization condition is valid regardless of the details of the materials in any spatial dimensions.

In this Letter, we give a partial solution to this problem by proving that the (longitudinal) spin susceptibility vanishes 
assuming only that excitations in the same total magnetization sector as the ground state are all gapped.  Although this statement might sound trivial at least empirically, we believe there is gap in logic in the existing literature.  Our result does not solve the problem completely, since it is, in principle, possible to realize a magnetization plateau even in the presence of gapless spinless excitations. In fact, we will see a concrete (trivial) example of such a situation.

We find it useful to formulate the problem in a slightly more abstract language.  Namely, we discuss a model with a U(1) symmetry $e^{i\theta \hat{Q}}$ in general and derive a non-perturbative constraint on the behavior of the U(1) charge $\langle\hat{Q}\rangle$ in response to an infinitesimal increase of the `chemical potential' $\mu$.  The charge $\hat{Q}$ and the field $\mu$ in the following discussion should be set $\hat{Q}=\sum_j(S-\hat{S}_j^z)$ and $\mu=B_z$ (the external magnetic field) for spin models.  Our general treatment gives us a coherent understanding on a similar phenomena in seemingly different setups but arising from the same mechanism, such as the `Mott plateau' in the Mott insulating phase of the Bose-Hubbard model~\cite{Zoller,Troyer}, or, more generally, the commensurate phases of the commensurate-incommensurate translations (see Ref.~\cite{cireview} and references therein).

\paragraph{Neutral and charged excitations.\,\,\,---}
To setup notations and clarify the addressed problem, let us review first that a finite energy gap of charge-neutral excitations does not immediately imply the vanishing charge susceptibility for interacting systems.  

Consider a Hamiltonian $\hat{H}$ with a U(1) symmetry $e^{i\theta \hat{Q}}$ defined on a $d$-dimensional lattice $\Lambda\subset\mathbb{Z}^d$.  
We assume that both the Hamiltonian $\hat{H}=\sum_{\vec{x}\in \Lambda}\hat{h}_{\vec{x}}$ and the U(1) charge $\hat{Q}=\sum_{\vec{x}\in \Lambda}\hat{n}_{\vec{x}}$ are the sum of the local operators, but we do not assume the translation symmetry.  Since $\hat{Q}$ commutes with $\hat{H}$, we decompose the total Hilbert space $\mathcal{H}$ into the direct sum of the charge $N$ sector, $\oplus_N\mathcal{H}_N$, and $\hat{H}$ can be diagonalized within each subspace $\mathcal{H}_N$.  The charge density in the ground state can be controlled by introducing an external field $\mu$ as $\hat{H}(\mu)\equiv \hat{H}-\mu \hat{Q}$.

Let $E_{0}^{N}(\mu)$ be the ground state energy of $\hat{H}(\mu)$ in the charge-$N$ sector, which may be degenerate.  For fixed $\Lambda$ and $\mu$, we denote by $N_\mu$ the maximum $N$ among those realize the minimum of $E_{0}^{N}(\mu)$:
\begin{equation}
N_\mu\equiv\max_{N}\{N\,|\,\text{$N$ minimizes $E_{0}^{N}(\mu)$}\}.
\end{equation}
It follows by definition that $N_{\mu'}\geq N_\mu$ when $\mu'>\mu$.  The static uniform charge susceptibility $\chi(\mu)$ at the zero temperature $T=0$ may be defined as
\begin{eqnarray}
\chi(\mu)&\equiv&\lim_{\mu'\searrow \mu}\frac{n(\mu')-n(\mu)}{\mu'-\mu}\notag\\
&=&\lim_{\mu'\searrow \mu}\lim_{V\rightarrow\infty}\frac{1}{V}\frac{N_{\mu'}-N_\mu}{\mu'-\mu}.\label{eq:chi1}
\end{eqnarray}
Here, $V$ is the volume of $\Lambda$ and $n(\mu)=\lim_{V\rightarrow\infty}\frac{N_\mu}{V}$ is the ground state charge density.  

\begin{figure}
\begin{center}
\includegraphics[width=0.9\columnwidth]{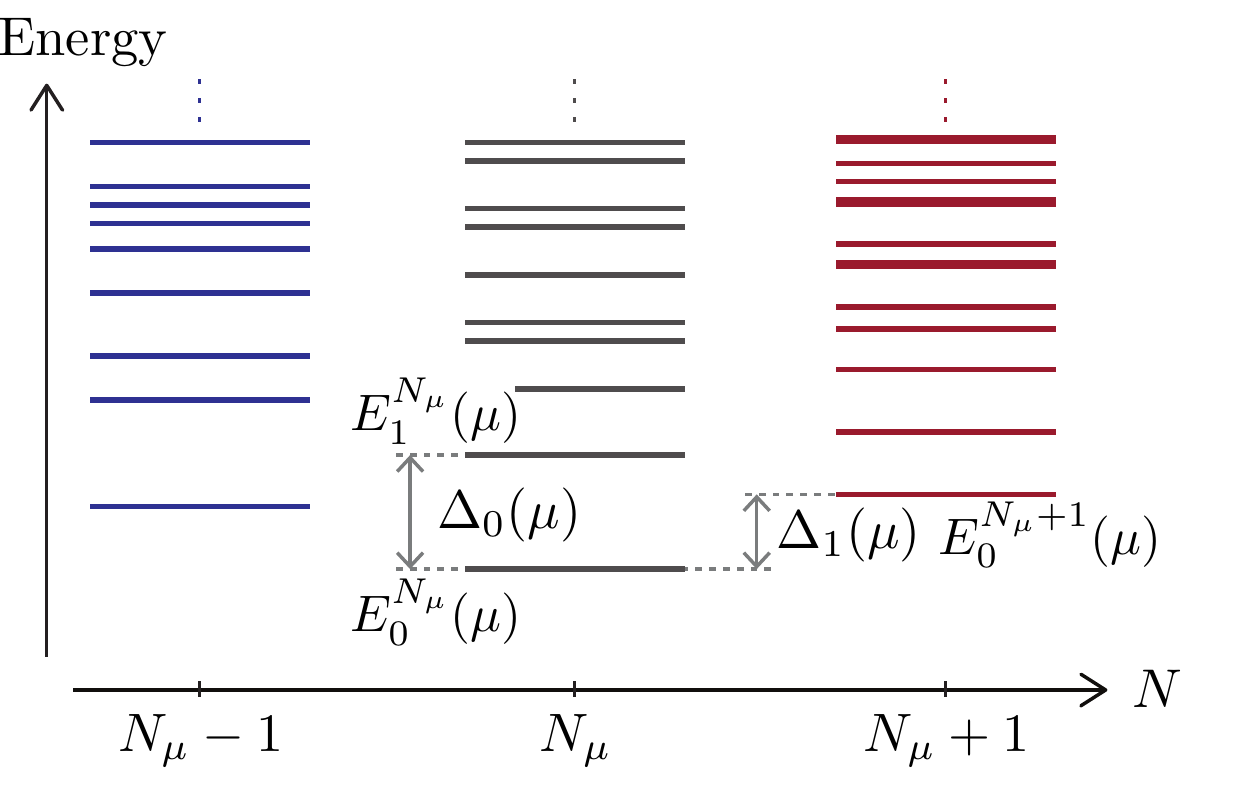}
\caption{Illustration of the neutral excitation gap $\Delta_0(\mu)$ and the charged excitation gap $\Delta_1(\mu)$.
\label{fig1}}
\end{center}
\end{figure}

It is widely believed that the charge susceptibility vanishes in the presence of a finite energy gap.  However, there are two completely different kinds of possible energy gaps and we need to distinguish them clearly (see Fig.~\ref{fig1}).  If $E_{1}^{N}(\mu)$ is the energy of the first excited state of $\hat{H}(\mu)$ in the charge-$N$ sector, the energy gap of charge-neutral excitations is given by
\begin{equation}
\Delta_0(\mu)\equiv\lim_{V\rightarrow\infty}\left[E_{1}^{N_\mu}(\mu)-E_{0}^{N_\mu}(\mu)\right].\label{eq:d0}
\end{equation}
On the other hand, the energy cost to add an extra charge to the ground state is
\begin{equation}
\Delta_1(\mu)\equiv\lim_{V\rightarrow\infty}\left[E_{0}^{N_\mu+1}(\mu)-E_{0}^{N_\mu}(\mu)\right].\label{eq:d1}
\end{equation}
When the charged excitations are gapped, $N_{\mu'}$ remains $N_\mu$ until $\mu'\,(\geq\mu)$ exceeds $\mu+\Delta_1(\mu)$. Thus, according to Eq.~\eqref{eq:chi1}, the charge susceptibility vanishes for this range of $\mu$ and the $n$-$\mu$ curve at $T=0$ exhibits a plateau.  Note that whether the neutral excitations are gapped or not is \emph{a priori} irrelevant to the existence of a plateau, since it is $\Delta_1(\mu)$ that matters. Yet, we will argue in the following that $\Delta_0(\mu)$ is still intimately related to the charge susceptibility.

The noninteracting fermionic systems are exceptional because the two gaps $\Delta_0(\mu)$ and $\Delta_1(\mu)$ are closely related.  Let $\{\epsilon_{n}\}_{n=1,2,3,\cdots}$ be the single-particle energy levels of $\hat{H}$ arranged in the increasing order $\epsilon_{1}\leq\epsilon_{2}\leq\epsilon_{3}\leq\cdots$.  By definition, $\mu$ must fall into the range $\epsilon_{N_\mu}\leq\mu\leq\epsilon_{N_\mu+1}$.  In this case, if the neutral excitation gap $\Delta_0(\mu)=\lim_{V\rightarrow\infty}(\epsilon_{N_\mu+1}-\epsilon_{N_\mu})$ is nonzero, then charged excitation gap $\Delta_1(\mu)=\lim_{V\rightarrow\infty}\epsilon_{N_\mu+1}-\mu$ is also nonzero and consequently the charge susceptibility vanishes for a finite range of $\mu$.  

In contrast, in strongly-correlated systems, $\Delta_0(\mu)$ and $\Delta_1(\mu)$ can, in principle, be independent and the charge susceptibility might be finite even when 
the neutral excitations are gapped.  What we will show in this Letter is that $\Delta_0(\mu)>0$, in fact, implies $\chi(\mu)=0$ regardless of the strength of the interactions and the dimensionality of the system.

\paragraph{Susceptibility in terms of correlation functions.\,\,\,---}
To this end it is more useful to express the charge susceptibility in terms of correlation functions.  Naively one applies an extra field $-\delta\mu \hat{Q}$ and computes the response of $\hat{Q}$ to the first order in $\delta\mu$. (Higher-oder terms in $\delta\mu$ do not contribute to the susceptibility.)  If we do so, however, the response trivially vanishes since $\hat{Q}$ commutes with $\hat{H}$ and is always conserved regardless of the energy gap.  The same issue frequently appears when computing the susceptibility of conserved quantities. For example, the non-analyticity of the Lindhard function at $\omega=k=0$~\cite{Altland,Mahan} has the same origin.

\begin{figure}
\begin{center}
\includegraphics[width=0.5\columnwidth]{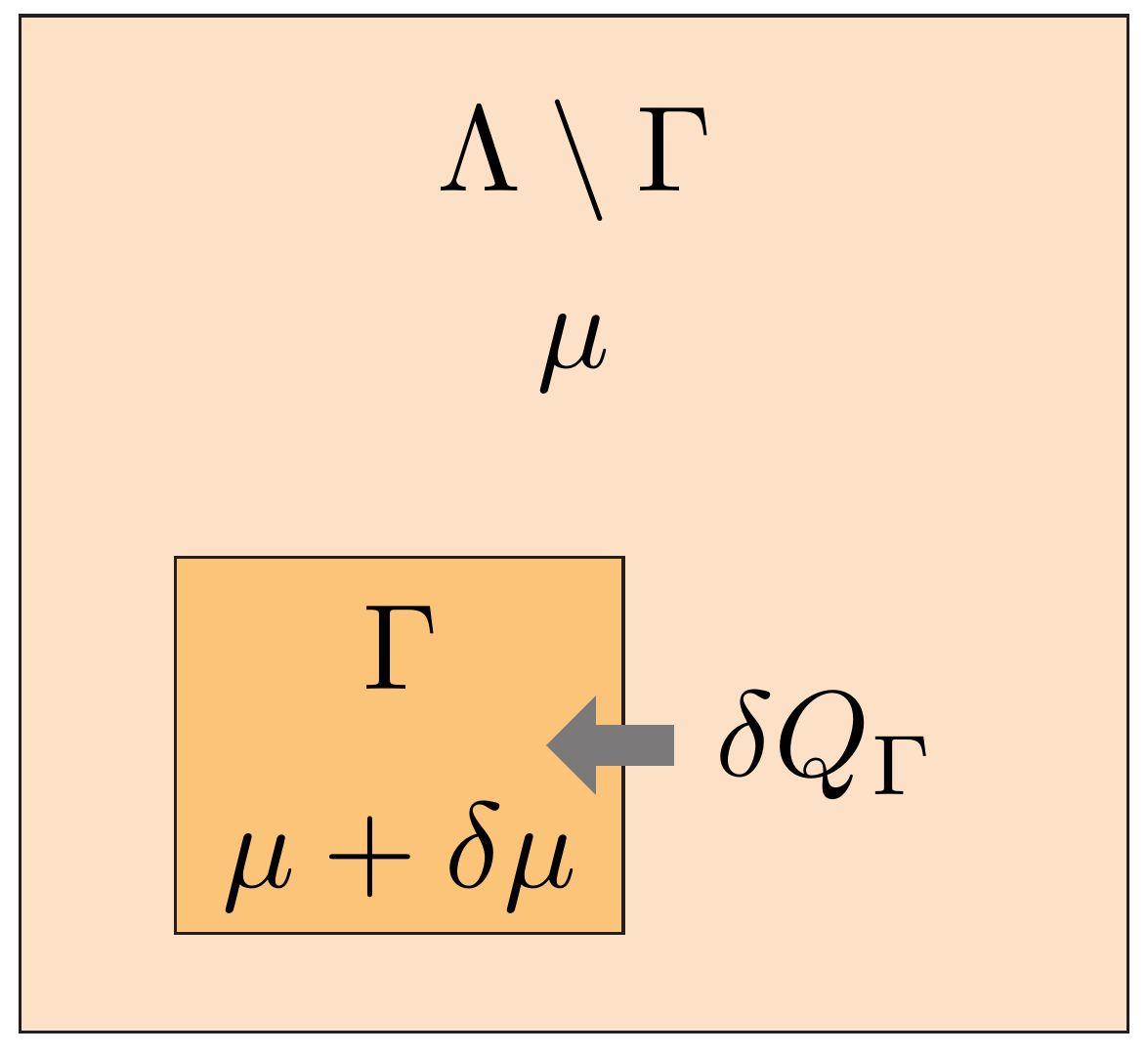}
\caption{The setup for computing the static charge susceptibility $\chi(\mu)$ from a correlation function.  The field $\mu$ is changed by $\delta \mu$ only in the subregion $\Gamma$ of the total system $\Lambda$. Consequently, an extra charge $\delta Q_\Gamma$ flows into $\Gamma$ from the ``charge reservoir" $\Lambda\setminus\Gamma$. \label{fig2}}
\end{center}
\end{figure}

To avoid this subtlety, let us take a sub-region $\Gamma$ in $\Lambda$ and imagine changing the external field $\mu$ only in this region, regarding the complement $\Lambda\setminus\Gamma$ a ``charge reservoir", as illustrated in Fig.~\ref{fig2}.  Even when we fix the total charge $\hat{Q}$ to be $N=N_\mu$ on $\Lambda$, an extra charge can flow into the subsystem $\Gamma$ from the reservoir $\Lambda\setminus\Gamma$ when the field $\mu$ is increased by $\delta\mu$ on $\Gamma$.  The region $\Gamma$ should be sufficiently large so that it by itself stands a thermodynamic system, and furthermore $\Lambda\setminus\Gamma$ should be much bigger than $\Gamma$ to be a reservoir.

For brevity we assume that the ground state of $\hat{H}(\mu)$ in the charge $N=N_\mu$ sector is unique, but we do not assume anything about $\Delta_1(\mu)$.  We will comment on a more general case of finite degeneracy later.  Let us denote the unique ground state by $|\Phi^{N_\mu}\rangle$ and introduce a shorthand notation $\langle O\rangle_\mu=\langle \Phi^{N_\mu}|O|\Phi^{N_\mu}\rangle$.  The change of $\hat{Q}_\Gamma\equiv\sum_{\vec{x}\in \Gamma}\hat{n}_{\vec{x}}$ in response to the perturbation $-\delta\mu \hat{Q}_\Gamma$ can be computed by the standard perturbation theory. We get
\begin{eqnarray}
\chi(\mu)&=&\lim_{V_\Gamma\rightarrow\infty}\lim_{V\rightarrow\infty}\frac{2}{V_\Gamma}\Big\langle \delta \hat{Q}_\Gamma \frac{1}{\hat{H}(\mu)-E_{0}^{N_\mu}(\mu)}\delta \hat{Q}_\Gamma \Big\rangle_\mu,
\label{eq:chi2}
\end{eqnarray}
where $\delta \hat{Q}_\Gamma\equiv \hat{Q}_\Gamma- \langle \hat{Q}_\Gamma\rangle_\mu$ and $V_\Gamma$ is the volume of $\Gamma$.  The same expression can be obtained alternatively as the static limit of Kubo's linear response function.   The ground state energy $E_{0}^{N_\mu}(\mu)$ decreases due to the perturbation. Equation~\eqref{eq:chi2} tells us that $\chi(\mu)$ is the second-derivative of the decrease of the ground state energy with respect to $\delta\mu$, divided by $V_\Gamma$.

Note that the intermediate states contributing to Eq.~\eqref{eq:chi2} have the same U(1) charge as the ground state as $\delta \hat{Q}_\Gamma$ commutes with $\hat{Q}$. This is why $\Delta_0(\mu)$ becomes relevant in the following discussion.  Assuming that $\Delta_0(\mu)\geq0$, we can derive the upper-bound of $\chi(\mu)$ as
\begin{eqnarray}
\chi(\mu)\leq\lim_{V_\Gamma\rightarrow\infty}\lim_{V\rightarrow\infty}\frac{2}{V_\Gamma}
\frac{\langle (\delta \hat{Q}_\Gamma)^2\rangle_\mu}{E_{1}^{N_\mu}(\mu)-E_{0}^{N_\mu}(\mu)}=\frac{2\,\sigma(\mu)}{\Delta_0(\mu)},\label{eq:finalchi}
\end{eqnarray}
where $\sigma(\mu)$ is the density fluctuation defined by
\begin{eqnarray}
\sigma(\mu)\equiv\lim_{V_\Gamma\rightarrow\infty}\lim_{V\rightarrow\infty}\frac{\langle(\delta \hat{Q}_\Gamma)^2\rangle_\mu}{V_\Gamma}.\label{eq:c}
\end{eqnarray}

\paragraph{The density fluctuation vanishes when $\Delta_0(\mu)>0$.\,\,\,---}
We are now going to prove that the density fluctuation $\sigma(\mu)$ vanishes when the neutral excitation gap $\Delta_0(\mu)$ is finite.  Given this proposition, one can immediately see from Eq.~\eqref{eq:finalchi} that $\chi(\mu)$ vanishes when $\Delta_0(\mu)>0$.  We will actually demonstrate the contraposition \,\,\,--- if $\sigma(\mu)\neq0$, there exists a low-energy state in the charge $N_\mu$ sector whose excitation energy is bounded by $\text{(const.)}\times L_\Gamma^{-1}$, where $L_\Gamma$ is the linear dimension of $\Gamma$, and thus $\Delta_0(\mu)=0$.  

The proof of this statement utilizes a trick using a `double commutator', introduced by Horsch and von~der~Linden~\cite{HorschvonderLinden,KomaTasaki}.  This technique was recently used in the context of quantum time crsytals~\cite{WatanabeOshikawa,Nayak}.  Let us introduce a variational state $|\Psi^{N_\mu}\rangle\equiv\frac{\delta \hat{Q}_\Gamma |\Phi^{N_\mu}\rangle}{\|\delta \hat{Q}_\Gamma |\Phi^{N_\mu}\rangle\|}$. It is well defined since $\|\delta \hat{Q}_\Gamma |\Phi^{N_\mu}\rangle\|^2=\langle(\delta \hat{Q}_\Gamma)^2\rangle_\mu\neq0$ [Eq.~\eqref{eq:c}] and fulfills the orthogonality condition $\langle \Phi^{N_\mu}|\Psi^{N_\mu}\rangle\propto\langle \delta \hat{Q}_\Gamma\rangle_\mu=0$.  The variational state $|\Psi^{N_\mu}\rangle$ and the ground state $|\Phi^{N_\mu}\rangle$  belong to the same sector of the U(1) charge, again because $[\hat{Q}, \hat{Q}_\Gamma]=0$.  Therefore, the neutral excitation gap $E_{1}^{N_\mu}(\mu)-E_{0}^{N_\mu}(\mu)$ can be bounded above by
\begin{eqnarray}
\langle \Psi^{N_\mu}|\hat{H}(\mu)|\Psi^{N_\mu}\rangle-E_{0}^{N_\mu}(\mu)=\frac{\langle[\delta\hat{Q}_\Gamma,[\hat{H},\delta\hat{Q}_\Gamma]]\rangle_\mu}{2\langle(\delta\hat{Q}_\Gamma)^2\rangle_\mu}.
\end{eqnarray}
Now we take the limit $V\rightarrow\infty$ first and ask how the numerator and the denominator behave in the limit of $V_\Gamma\rightarrow\infty$.  We observe that the support of the commutator $[\hat{H},\delta \hat{Q}_\Gamma]$ is only near the boundary $\partial\Gamma$ of $\Gamma$ because of the assumed U(1) symmetry and the locality of the Hamiltonian.  Thus the numerator is at most the order of $V_{\partial\Gamma}$ (the volume of the boundary of $\Gamma$), while the denominator grows with $V_\Gamma$ since $\sigma(\mu)>0$.  Therefore, the neutral excitation gap is bounded above by $O(V_{\partial\Gamma}/V_\Gamma)=O(L_\Gamma^{-1})$ and vanishes in the limit of $V_\Gamma\rightarrow \infty$. 

One might think that our argument resembles the original argument of the Lieb-Schultz-Mattis theorem~\cite{LiebSchultzMattis} in that a variational state, which is shown to be orthogonal to the ground state and have a vanishingly small excitation energy, is constructed out of the ground state by applying an operator. However, the argument is, in fact, completely different as one can see from the fact that we did not assume any translation symmetry but instead assumed $\sigma(\mu)>0$.

We would hasten to emphasize that the above result does not trivially follow from the well-known fact that the truncated correlation function of a gapped system decays exponentially with the distance~\cite{Hastings,HastingsKoma}.  Also, since we take the limit $V\rightarrow\infty$ before taking $V_\Gamma\rightarrow\infty$, one cannot argue $\lim_{V_\Gamma\rightarrow\infty}\lim_{V\rightarrow\infty}\langle(\delta \hat{Q}_\Gamma)^2\rangle_\mu=0$ simply because $|\Phi^{N_\mu}\rangle$ is an eigenstate of $\hat{Q}$. (It would be the case for $\lim_{V\rightarrow\infty}\lim_{\Gamma\rightarrow\Lambda}\langle(\delta \hat{Q}_\Gamma)^2\rangle_\mu=0$).  To support this point, we did a simple exercise for the 1D tight-binding model $\hat{H}=-t\sum_{x}(\hat{c}_{x+1}^\dagger \hat{c}_x+\text{h.c.})$ with the periodic boundary condition.  We found $\lim_{V\rightarrow\infty}\langle(\delta \hat{Q}_\Gamma)^2\rangle_\mu$ with $\delta \hat{Q}_\Gamma=\sum_{x\in\Gamma}(\hat{c}_x^\dagger \hat{c}_x-n)$ is nonzero and exhibits a logarithmic divergence $\frac{1}{\pi^2}\log V_\Gamma$ when $0<n<1$, i.e., in the presence of a Fermi surface.  

The above trick is also useful to estimate the magnitude of the charge fluctuation even in gapless phases.  Using the Schwartz inequality $|\langle A|B\rangle|^2\leq \langle A|A\rangle\langle B|B\rangle$ for two states $|A\rangle=\left[\hat{H}(\mu)-E_{0}^{N_\mu}(\mu)\right]^{-1/2}|\Phi^{N_\mu}\rangle$ and $|B\rangle=\left[\hat{H}(\mu)-E_{0}^{N_\mu}(\mu)\right]^{1/2}|\Phi^{N_\mu}\rangle$, we have
\begin{eqnarray}
\Big|\frac{\langle(\delta \hat{Q}_\Gamma)^2\rangle_\mu}{V_\Gamma}\Big|^2&\leq&
\frac{2}{V_\Gamma}\Big\langle \delta \hat{Q}_\Gamma \frac{1}{\hat{H}(\mu)-E_{0}^{N_\mu}(\mu)}\delta \hat{Q}_\Gamma \Big\rangle_\mu\notag\\
&&\times\frac{1}{4V_{\Gamma}}\langle[\delta\hat{Q}_\Gamma,[\hat{H},\delta\hat{Q}_\Gamma]]\rangle_\mu.
\end{eqnarray}
In the large volume limit, the left-hand side is the square of the density fluctuation $\sigma(\mu)$, while the first line of the right-hand side is the charge susceptibility $\chi(\mu)$ and the second line is $O(V_{\partial\Gamma}/V_{\Gamma})$ as discussed above. Therefore, $\sigma(\mu)$ vanishes when $\chi(\mu)$ is finite; it can be nonzero at $T=0$ only when $\chi(\mu)$ diverges. For instance, in the 1D tight-binding model discussed above, $\sigma(\mu)=\lim_{V_\Gamma\rightarrow\infty}\frac{1}{\pi^2}\frac{\log V_\Gamma}{V_\Gamma}$ indeed vanishes because $\chi(\mu)=\frac{1}{2\pi t\sin(\pi n)}$ is finite for $0<n<1$.

\paragraph{Examples of $\chi(\mu)=0$ when $\Delta_0(\mu)=0$.\,\,\,---}
We have proved that the static charge susceptibility $\chi(\mu)$ vanishes at $T=0$ when the neutral excitations are gapped.  Although the converse sounds plausible, it does not hold in general. Since known counterexamples~\cite{Frahm,Cabra2000,Cabra2002,Roux,Lamas} are somewhat complicated, let us discuss here a simple example of noninteracting band semimetals.  
For noninteracting electrons, $\chi(\mu)$ coincides with the density of states $D(\mu)$ per unit volume and thus $\chi(\mu)$ vanishes when the dimensionality of the Fermi surface reduces, for example, at Dirac or Weyl points in 3D. Nonetheless, there are both neutral and charged gapless excitations. Hence, in general, $\chi(\mu)=0$ does not imply $\Delta_0(\mu)>0$ or $\Delta_1(\mu)>0$.

Even if one assumes a plateau, not only $\chi(\mu)=0$ at a single point of $\mu$, there is still an example with $\Delta_0(\mu)=0$.
Imagine applying a strong magnetic magnetic field to a system of spin-$1/2$ electrons in 1D with the translation invariance.  Assume that the translation symmetry is not broken spontaneously.  We set the total number of electrons per unit cell to be, say, $1/2$ so that the system is gapless due to the Lieb-Shultz-Mattis theorem~\cite{LiebSchultzMattis,Affleck1986,Yamanaka,HastingsLSM,Sid,PNAS}. If the magnetic field is strong enough, the spin of electrons will be completely polarized and spinful excitations will have a gap comparable to the field, i.e., $\Delta_1(\mu)>0$. As a result, there will be a magnetization plateau, regardless of gapless neutral (spinless) excitations, $\Delta_0(\mu)=0$.

\begin{figure}
\begin{center}
\includegraphics[width=0.5\columnwidth]{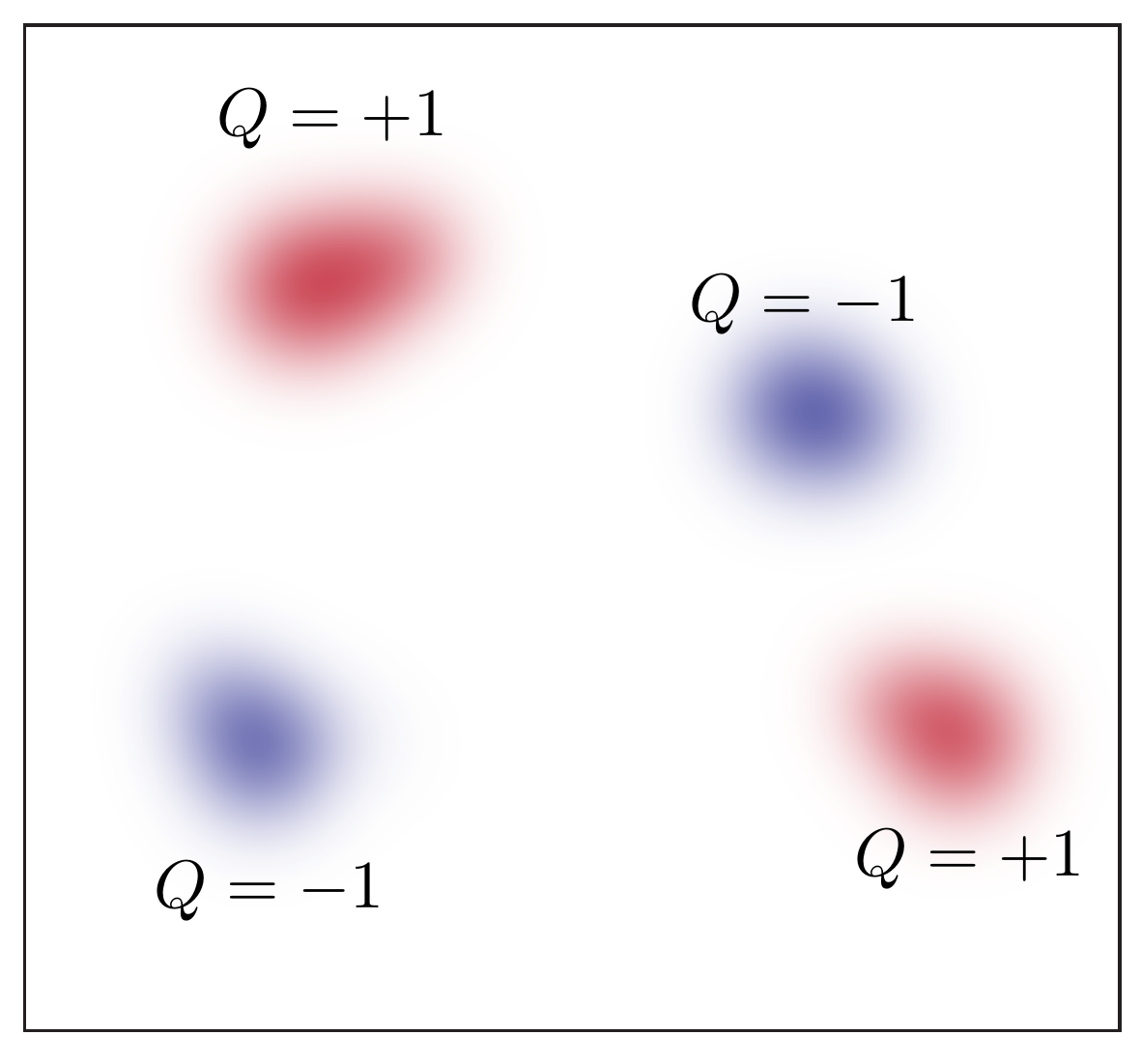}
\caption{When the quasi-particle picture is valid, one can construct gapless \emph{neutral} excitations using gapless \emph{charged} excitations, assuming that both positively and negatively charged gapless excitations exist and can be localized. \label{fig3}}
\end{center}
\end{figure}

\paragraph{Discussions.\,\,\,---}
Our result is consistent with the quasi-particle description of low-energy excitations in many-body systems. If the susceptibility is continuous and nonzero around $\mu$, there must be both positively- and negatively-charged gapless excitations.  If these excitations are particle-like, one can readily construct gapless neutral excitations by distributing an arbitrary number of `particles' and the same number of `anti-particles' far away with each other so that the interaction among them can be neglected (Fig.~\ref{fig3}). Therefore, a finite continuous susceptibility around $\mu$ implies that both $\Delta_0(\mu)$ and $\Delta_1(\mu)$ vanish.  We have shown the same statement without relying on the quasi-partible picture.

In the proof we assumed the uniqueness of the ground state in the charge-$N_\mu$ sector.  However, this assumption might be violated in a magnetization plateau accompanying spontaneous breaking of translation symmetry.  For example, $1/3$ magnetization plateau appears in triangular lattice systems when the unit cell is enlarged by three times so that $S-m_z$ per the new unit cell becomes an integer~\cite{Richter,Miyashita}.  Consequently there will be three degenerate ground states.  When the ground states are degenerate, normally we have to use the degenerate perturbation theory.  However, it is not the case when the degeneracy originates from spontaneously symmetry breaking. To see this, let $|\Phi_m^{N_\mu}\rangle$ ($m=1,2,\ldots,M^{N_\mu}$) be the (quasi-)degenerate ground states.   In this case, the matrix elements $\langle 
\Phi_m^{N_\mu}|\delta \hat{Q}_\Gamma|\Phi_n^{N_\mu}\rangle$ ($m\neq n$) are exponentially small in the system size $e^{-cV}$ and the degenerate perturbation theory automatically reduces to the non-degenerate one.

We proved only that $\chi(\mu)$ vanishes at a single value of $\mu$ assuming that the neutral excitations are gapped at this $\mu$.  
This alone does not necessarily mean that there is a plateau around $\mu$.  It would be an interesting future work to ask when the spectrum has a stability against an infinitesimal change of $\mu$ and whether $\Delta_0(\mu)>0$ implies a plateau in general or not.

\begin{acknowledgments}
I wish to thank Tohru Koma for bringing the present issue to my attention and also for very helpful discussions.  I also thank Masaki Oshikawa, Shunsuke Furukawa, Hoi Chun Po, and Hal Tasaki for useful comments and discussions.
\end{acknowledgments}

\bibliography{references}

\begin{thebibliography}{44}%
\makeatletter
\providecommand \@ifxundefined [1]{%
 \@ifx{#1\undefined}
}%
\providecommand \@ifnum [1]{%
 \ifnum #1\expandafter \@firstoftwo
 \else \expandafter \@secondoftwo
 \fi
}%
\providecommand \@ifx [1]{%
 \ifx #1\expandafter \@firstoftwo
 \else \expandafter \@secondoftwo
 \fi
}%
\providecommand \natexlab [1]{#1}%
\providecommand \enquote  [1]{``#1''}%
\providecommand \bibnamefont  [1]{#1}%
\providecommand \bibfnamefont [1]{#1}%
\providecommand \citenamefont [1]{#1}%
\providecommand \href@noop [0]{\@secondoftwo}%
\providecommand \href [0]{\begingroup \@sanitize@url \@href}%
\providecommand \@href[1]{\@@startlink{#1}\@@href}%
\providecommand \@@href[1]{\endgroup#1\@@endlink}%
\providecommand \@sanitize@url [0]{\catcode `\\12\catcode `\$12\catcode
  `\&12\catcode `\#12\catcode `\^12\catcode `\_12\catcode `\%12\relax}%
\providecommand \@@startlink[1]{}%
\providecommand \@@endlink[0]{}%
\providecommand \url  [0]{\begingroup\@sanitize@url \@url }%
\providecommand \@url [1]{\endgroup\@href {#1}{\urlprefix }}%
\providecommand \urlprefix  [0]{URL }%
\providecommand \Eprint [0]{\href }%
\providecommand \doibase [0]{http://dx.doi.org/}%
\providecommand \selectlanguage [0]{\@gobble}%
\providecommand \bibinfo  [0]{\@secondoftwo}%
\providecommand \bibfield  [0]{\@secondoftwo}%
\providecommand \translation [1]{[#1]}%
\providecommand \BibitemOpen [0]{}%
\providecommand \bibitemStop [0]{}%
\providecommand \bibitemNoStop [0]{.\EOS\space}%
\providecommand \EOS [0]{\spacefactor3000\relax}%
\providecommand \BibitemShut  [1]{\csname bibitem#1\endcsname}%
\let\auto@bib@innerbib\@empty
\bibitem [{\citenamefont {Narumi}\ \emph {et~al.}(1998)\citenamefont {Narumi},
  \citenamefont {Hagiwara}, \citenamefont {Sato}, \citenamefont {Kindo},
  \citenamefont {Nakano},\ and\ \citenamefont {Takahashi}}]{Narumi}%
  \BibitemOpen
  \bibfield  {author} {\bibinfo {author} {\bibfnamefont {Y.}~\bibnamefont
  {Narumi}}, \bibinfo {author} {\bibfnamefont {M.}~\bibnamefont {Hagiwara}},
  \bibinfo {author} {\bibfnamefont {R.}~\bibnamefont {Sato}}, \bibinfo {author}
  {\bibfnamefont {K.}~\bibnamefont {Kindo}}, \bibinfo {author} {\bibfnamefont
  {H.}~\bibnamefont {Nakano}}, \ and\ \bibinfo {author} {\bibfnamefont
  {M.}~\bibnamefont {Takahashi}},\ }\href {\doibase
  http://dx.doi.org/10.1016/S0921-4526(97)00974-5} {\bibfield  {journal}
  {\bibinfo  {journal} {Physica B: Cond. Mat.}\ }\textbf {\bibinfo {volume}
  {246--247}},\ \bibinfo {pages} {509} (\bibinfo {year} {1998})}\BibitemShut
  {NoStop}%
\bibitem [{\citenamefont {Kikuchi}\ \emph {et~al.}(2005)\citenamefont
  {Kikuchi}, \citenamefont {Fujii}, \citenamefont {Chiba}, \citenamefont
  {Mitsudo}, \citenamefont {Idehara}, \citenamefont {Tonegawa}, \citenamefont
  {Okamoto}, \citenamefont {Sakai}, \citenamefont {Kuwai},\ and\ \citenamefont
  {Ohta}}]{Kikuchi}%
  \BibitemOpen
  \bibfield  {author} {\bibinfo {author} {\bibfnamefont {H.}~\bibnamefont
  {Kikuchi}}, \bibinfo {author} {\bibfnamefont {Y.}~\bibnamefont {Fujii}},
  \bibinfo {author} {\bibfnamefont {M.}~\bibnamefont {Chiba}}, \bibinfo
  {author} {\bibfnamefont {S.}~\bibnamefont {Mitsudo}}, \bibinfo {author}
  {\bibfnamefont {T.}~\bibnamefont {Idehara}}, \bibinfo {author} {\bibfnamefont
  {T.}~\bibnamefont {Tonegawa}}, \bibinfo {author} {\bibfnamefont
  {K.}~\bibnamefont {Okamoto}}, \bibinfo {author} {\bibfnamefont
  {T.}~\bibnamefont {Sakai}}, \bibinfo {author} {\bibfnamefont
  {T.}~\bibnamefont {Kuwai}}, \ and\ \bibinfo {author} {\bibfnamefont
  {H.}~\bibnamefont {Ohta}},\ }\href {\doibase 10.1103/PhysRevLett.94.227201}
  {\bibfield  {journal} {\bibinfo  {journal} {Phys. Rev. Lett.}\ }\textbf
  {\bibinfo {volume} {94}},\ \bibinfo {pages} {227201} (\bibinfo {year}
  {2005})}\BibitemShut {NoStop}%
\bibitem [{\citenamefont {He}\ \emph {et~al.}(2009)\citenamefont {He},
  \citenamefont {Yamaura}, \citenamefont {Ueda},\ and\ \citenamefont
  {Cheng}}]{Zhangzhen}%
  \BibitemOpen
  \bibfield  {author} {\bibinfo {author} {\bibfnamefont {Z.}~\bibnamefont
  {He}}, \bibinfo {author} {\bibfnamefont {J.-I.}\ \bibnamefont {Yamaura}},
  \bibinfo {author} {\bibfnamefont {Y.}~\bibnamefont {Ueda}}, \ and\ \bibinfo
  {author} {\bibfnamefont {W.}~\bibnamefont {Cheng}},\ }\href
  {http://dx.doi.org/10.1021/ja902623b} {\bibfield  {journal} {\bibinfo
  {journal} {J. Am. Chem. Soc.}\ }\textbf {\bibinfo {volume} {131}},\ \bibinfo
  {pages} {7554} (\bibinfo {year} {2009})}\BibitemShut {NoStop}%
\bibitem [{\citenamefont {Hase}\ \emph {et~al.}(2006)\citenamefont {Hase},
  \citenamefont {Kohno}, \citenamefont {Kitazawa}, \citenamefont {Tsujii},
  \citenamefont {Suzuki}, \citenamefont {Ozawa}, \citenamefont {Kido},
  \citenamefont {Imai},\ and\ \citenamefont {Hu}}]{Hase}%
  \BibitemOpen
  \bibfield  {author} {\bibinfo {author} {\bibfnamefont {M.}~\bibnamefont
  {Hase}}, \bibinfo {author} {\bibfnamefont {M.}~\bibnamefont {Kohno}},
  \bibinfo {author} {\bibfnamefont {H.}~\bibnamefont {Kitazawa}}, \bibinfo
  {author} {\bibfnamefont {N.}~\bibnamefont {Tsujii}}, \bibinfo {author}
  {\bibfnamefont {O.}~\bibnamefont {Suzuki}}, \bibinfo {author} {\bibfnamefont
  {K.}~\bibnamefont {Ozawa}}, \bibinfo {author} {\bibfnamefont
  {G.}~\bibnamefont {Kido}}, \bibinfo {author} {\bibfnamefont {M.}~\bibnamefont
  {Imai}}, \ and\ \bibinfo {author} {\bibfnamefont {X.}~\bibnamefont {Hu}},\
  }\href {\doibase 10.1103/PhysRevB.73.104419} {\bibfield  {journal} {\bibinfo
  {journal} {Phys. Rev. B}\ }\textbf {\bibinfo {volume} {73}},\ \bibinfo
  {pages} {104419} (\bibinfo {year} {2006})}\BibitemShut {NoStop}%
\bibitem [{\citenamefont {Kageyama}\ \emph {et~al.}(1999)\citenamefont
  {Kageyama}, \citenamefont {Yoshimura}, \citenamefont {Stern}, \citenamefont
  {Mushnikov}, \citenamefont {Onizuka}, \citenamefont {Kato}, \citenamefont
  {Kosuge}, \citenamefont {Slichter}, \citenamefont {Goto},\ and\ \citenamefont
  {Ueda}}]{Kageyama}%
  \BibitemOpen
  \bibfield  {author} {\bibinfo {author} {\bibfnamefont {H.}~\bibnamefont
  {Kageyama}}, \bibinfo {author} {\bibfnamefont {K.}~\bibnamefont {Yoshimura}},
  \bibinfo {author} {\bibfnamefont {R.}~\bibnamefont {Stern}}, \bibinfo
  {author} {\bibfnamefont {N.~V.}\ \bibnamefont {Mushnikov}}, \bibinfo {author}
  {\bibfnamefont {K.}~\bibnamefont {Onizuka}}, \bibinfo {author} {\bibfnamefont
  {M.}~\bibnamefont {Kato}}, \bibinfo {author} {\bibfnamefont {K.}~\bibnamefont
  {Kosuge}}, \bibinfo {author} {\bibfnamefont {C.~P.}\ \bibnamefont
  {Slichter}}, \bibinfo {author} {\bibfnamefont {T.}~\bibnamefont {Goto}}, \
  and\ \bibinfo {author} {\bibfnamefont {Y.}~\bibnamefont {Ueda}},\ }\href
  {\doibase 10.1103/PhysRevLett.82.3168} {\bibfield  {journal} {\bibinfo
  {journal} {Phys. Rev. Lett.}\ }\textbf {\bibinfo {volume} {82}},\ \bibinfo
  {pages} {3168} (\bibinfo {year} {1999})}\BibitemShut {NoStop}%
\bibitem [{\citenamefont {Onizuka}\ \emph {et~al.}(2000)\citenamefont
  {Onizuka}, \citenamefont {Kageyama}, \citenamefont {Narumi}, \citenamefont
  {Kindo}, \citenamefont {Ueda},\ and\ \citenamefont {Goto}}]{Onizuka}%
  \BibitemOpen
  \bibfield  {author} {\bibinfo {author} {\bibfnamefont {K.}~\bibnamefont
  {Onizuka}}, \bibinfo {author} {\bibfnamefont {H.}~\bibnamefont {Kageyama}},
  \bibinfo {author} {\bibfnamefont {Y.}~\bibnamefont {Narumi}}, \bibinfo
  {author} {\bibfnamefont {K.}~\bibnamefont {Kindo}}, \bibinfo {author}
  {\bibfnamefont {Y.}~\bibnamefont {Ueda}}, \ and\ \bibinfo {author}
  {\bibfnamefont {T.}~\bibnamefont {Goto}},\ }\href {\doibase
  10.1143/JPSJ.69.1016} {\bibfield  {journal} {\bibinfo  {journal} {J. Phys.
  Soc. Jpn.}\ }\textbf {\bibinfo {volume} {69}},\ \bibinfo {pages} {1016}
  (\bibinfo {year} {2000})}\BibitemShut {NoStop}%
\bibitem [{\citenamefont {Ono}\ \emph {et~al.}(2003)\citenamefont {Ono},
  \citenamefont {Tanaka}, \citenamefont {Aruga~Katori}, \citenamefont
  {Ishikawa}, \citenamefont {Mitamura},\ and\ \citenamefont {Goto}}]{Ono}%
  \BibitemOpen
  \bibfield  {author} {\bibinfo {author} {\bibfnamefont {T.}~\bibnamefont
  {Ono}}, \bibinfo {author} {\bibfnamefont {H.}~\bibnamefont {Tanaka}},
  \bibinfo {author} {\bibfnamefont {H.}~\bibnamefont {Aruga~Katori}}, \bibinfo
  {author} {\bibfnamefont {F.}~\bibnamefont {Ishikawa}}, \bibinfo {author}
  {\bibfnamefont {H.}~\bibnamefont {Mitamura}}, \ and\ \bibinfo {author}
  {\bibfnamefont {T.}~\bibnamefont {Goto}},\ }\href {\doibase
  10.1103/PhysRevB.67.104431} {\bibfield  {journal} {\bibinfo  {journal} {Phys.
  Rev. B}\ }\textbf {\bibinfo {volume} {67}},\ \bibinfo {pages} {104431}
  (\bibinfo {year} {2003})}\BibitemShut {NoStop}%
\bibitem [{\citenamefont {Shirata}\ \emph {et~al.}(2012)\citenamefont
  {Shirata}, \citenamefont {Tanaka}, \citenamefont {Matsuo},\ and\
  \citenamefont {Kindo}}]{Shirata}%
  \BibitemOpen
  \bibfield  {author} {\bibinfo {author} {\bibfnamefont {Y.}~\bibnamefont
  {Shirata}}, \bibinfo {author} {\bibfnamefont {H.}~\bibnamefont {Tanaka}},
  \bibinfo {author} {\bibfnamefont {A.}~\bibnamefont {Matsuo}}, \ and\ \bibinfo
  {author} {\bibfnamefont {K.}~\bibnamefont {Kindo}},\ }\href {\doibase
  10.1103/PhysRevLett.108.057205} {\bibfield  {journal} {\bibinfo  {journal}
  {Phys. Rev. Lett.}\ }\textbf {\bibinfo {volume} {108}},\ \bibinfo {pages}
  {057205} (\bibinfo {year} {2012})}\BibitemShut {NoStop}%
\bibitem [{\citenamefont {Ishikawa}\ \emph {et~al.}(2015)\citenamefont
  {Ishikawa}, \citenamefont {Yoshida}, \citenamefont {Nawa}, \citenamefont
  {Jeong}, \citenamefont {Kr\"amer}, \citenamefont
  {Horvati\ifmmode~\acute{c}\else \'{c}\fi{}}, \citenamefont {Berthier},
  \citenamefont {Takigawa}, \citenamefont {Akaki}, \citenamefont {Miyake},
  \citenamefont {Tokunaga}, \citenamefont {Kindo}, \citenamefont {Yamaura},
  \citenamefont {Okamoto},\ and\ \citenamefont {Hiroi}}]{Ishikawa}%
  \BibitemOpen
  \bibfield  {author} {\bibinfo {author} {\bibfnamefont {H.}~\bibnamefont
  {Ishikawa}}, \bibinfo {author} {\bibfnamefont {M.}~\bibnamefont {Yoshida}},
  \bibinfo {author} {\bibfnamefont {K.}~\bibnamefont {Nawa}}, \bibinfo {author}
  {\bibfnamefont {M.}~\bibnamefont {Jeong}}, \bibinfo {author} {\bibfnamefont
  {S.}~\bibnamefont {Kr\"amer}}, \bibinfo {author} {\bibfnamefont
  {M.}~\bibnamefont {Horvati\ifmmode~\acute{c}\else \'{c}\fi{}}}, \bibinfo
  {author} {\bibfnamefont {C.}~\bibnamefont {Berthier}}, \bibinfo {author}
  {\bibfnamefont {M.}~\bibnamefont {Takigawa}}, \bibinfo {author}
  {\bibfnamefont {M.}~\bibnamefont {Akaki}}, \bibinfo {author} {\bibfnamefont
  {A.}~\bibnamefont {Miyake}}, \bibinfo {author} {\bibfnamefont
  {M.}~\bibnamefont {Tokunaga}}, \bibinfo {author} {\bibfnamefont
  {K.}~\bibnamefont {Kindo}}, \bibinfo {author} {\bibfnamefont
  {J.}~\bibnamefont {Yamaura}}, \bibinfo {author} {\bibfnamefont
  {Y.}~\bibnamefont {Okamoto}}, \ and\ \bibinfo {author} {\bibfnamefont
  {Z.}~\bibnamefont {Hiroi}},\ }\href {\doibase 10.1103/PhysRevLett.114.227202}
  {\bibfield  {journal} {\bibinfo  {journal} {Phys. Rev. Lett.}\ }\textbf
  {\bibinfo {volume} {114}},\ \bibinfo {pages} {227202} (\bibinfo {year}
  {2015})}\BibitemShut {NoStop}%
\bibitem [{\citenamefont {Shiramura}\ \emph {et~al.}(1998)\citenamefont
  {Shiramura}, \citenamefont {Takatsu}, \citenamefont {Kurniawan},
  \citenamefont {Tanaka}, \citenamefont {Uekusa}, \citenamefont {Ohashi},
  \citenamefont {Takizawa}, \citenamefont {Mitamura},\ and\ \citenamefont
  {Goto}}]{Shiramura}%
  \BibitemOpen
  \bibfield  {author} {\bibinfo {author} {\bibfnamefont {W.}~\bibnamefont
  {Shiramura}}, \bibinfo {author} {\bibfnamefont {K.-i.}\ \bibnamefont
  {Takatsu}}, \bibinfo {author} {\bibfnamefont {B.}~\bibnamefont {Kurniawan}},
  \bibinfo {author} {\bibfnamefont {H.}~\bibnamefont {Tanaka}}, \bibinfo
  {author} {\bibfnamefont {H.}~\bibnamefont {Uekusa}}, \bibinfo {author}
  {\bibfnamefont {Y.}~\bibnamefont {Ohashi}}, \bibinfo {author} {\bibfnamefont
  {K.}~\bibnamefont {Takizawa}}, \bibinfo {author} {\bibfnamefont
  {H.}~\bibnamefont {Mitamura}}, \ and\ \bibinfo {author} {\bibfnamefont
  {T.}~\bibnamefont {Goto}},\ }\href {\doibase 10.1143/JPSJ.67.1548} {\bibfield
   {journal} {\bibinfo  {journal} {J. Phys. Soc. Jpn.}\ }\textbf {\bibinfo
  {volume} {67}},\ \bibinfo {pages} {1548} (\bibinfo {year}
  {1998})}\BibitemShut {NoStop}%
\bibitem [{\citenamefont {Ueda}\ \emph {et~al.}(2005)\citenamefont {Ueda},
  \citenamefont {Katori}, \citenamefont {Mitamura}, \citenamefont {Goto},\ and\
  \citenamefont {Takagi}}]{Ueda}%
  \BibitemOpen
  \bibfield  {author} {\bibinfo {author} {\bibfnamefont {H.}~\bibnamefont
  {Ueda}}, \bibinfo {author} {\bibfnamefont {H.~A.}\ \bibnamefont {Katori}},
  \bibinfo {author} {\bibfnamefont {H.}~\bibnamefont {Mitamura}}, \bibinfo
  {author} {\bibfnamefont {T.}~\bibnamefont {Goto}}, \ and\ \bibinfo {author}
  {\bibfnamefont {H.}~\bibnamefont {Takagi}},\ }\href {\doibase
  10.1103/PhysRevLett.94.047202} {\bibfield  {journal} {\bibinfo  {journal}
  {Phys. Rev. Lett.}\ }\textbf {\bibinfo {volume} {94}},\ \bibinfo {pages}
  {047202} (\bibinfo {year} {2005})}\BibitemShut {NoStop}%
\bibitem [{\citenamefont {Ueda}\ \emph {et~al.}(2006)\citenamefont {Ueda},
  \citenamefont {Mitamura}, \citenamefont {Goto},\ and\ \citenamefont
  {Ueda}}]{Ueda2}%
  \BibitemOpen
  \bibfield  {author} {\bibinfo {author} {\bibfnamefont {H.}~\bibnamefont
  {Ueda}}, \bibinfo {author} {\bibfnamefont {H.}~\bibnamefont {Mitamura}},
  \bibinfo {author} {\bibfnamefont {T.}~\bibnamefont {Goto}}, \ and\ \bibinfo
  {author} {\bibfnamefont {Y.}~\bibnamefont {Ueda}},\ }\href {\doibase
  10.1103/PhysRevB.73.094415} {\bibfield  {journal} {\bibinfo  {journal} {Phys.
  Rev. B}\ }\textbf {\bibinfo {volume} {73}},\ \bibinfo {pages} {094415}
  (\bibinfo {year} {2006})}\BibitemShut {NoStop}%
\bibitem [{\citenamefont {Oshikawa}\ \emph {et~al.}(1997)\citenamefont
  {Oshikawa}, \citenamefont {Yamanaka},\ and\ \citenamefont
  {Affleck}}]{OshikawaYamanakaAffleck}%
  \BibitemOpen
  \bibfield  {author} {\bibinfo {author} {\bibfnamefont {M.}~\bibnamefont
  {Oshikawa}}, \bibinfo {author} {\bibfnamefont {M.}~\bibnamefont {Yamanaka}},
  \ and\ \bibinfo {author} {\bibfnamefont {I.}~\bibnamefont {Affleck}},\ }\href
  {\doibase 10.1103/PhysRevLett.78.1984} {\bibfield  {journal} {\bibinfo
  {journal} {Phys. Rev. Lett.}\ }\textbf {\bibinfo {volume} {78}},\ \bibinfo
  {pages} {1984} (\bibinfo {year} {1997})}\BibitemShut {NoStop}%
\bibitem [{\citenamefont {Fledderjohann}\ \emph {et~al.}(1999)\citenamefont
  {Fledderjohann}, \citenamefont {Gerhardt}, \citenamefont {Karbach},
  \citenamefont {M\"utter},\ and\ \citenamefont {Wie\ss{}ner}}]{Fledderjohann}%
  \BibitemOpen
  \bibfield  {author} {\bibinfo {author} {\bibfnamefont {A.}~\bibnamefont
  {Fledderjohann}}, \bibinfo {author} {\bibfnamefont {C.}~\bibnamefont
  {Gerhardt}}, \bibinfo {author} {\bibfnamefont {M.}~\bibnamefont {Karbach}},
  \bibinfo {author} {\bibfnamefont {K.-H.}\ \bibnamefont {M\"utter}}, \ and\
  \bibinfo {author} {\bibfnamefont {R.}~\bibnamefont {Wie\ss{}ner}},\ }\href
  {\doibase 10.1103/PhysRevB.59.991} {\bibfield  {journal} {\bibinfo  {journal}
  {Phys. Rev. B}\ }\textbf {\bibinfo {volume} {59}},\ \bibinfo {pages} {991}
  (\bibinfo {year} {1999})}\BibitemShut {NoStop}%
\bibitem [{\citenamefont {Oshikawa}(2000)}]{Oshikawa}%
  \BibitemOpen
  \bibfield  {author} {\bibinfo {author} {\bibfnamefont {M.}~\bibnamefont
  {Oshikawa}},\ }\href {\doibase 10.1103/PhysRevLett.84.1535} {\bibfield
  {journal} {\bibinfo  {journal} {Phys. Rev. Lett.}\ }\textbf {\bibinfo
  {volume} {84}},\ \bibinfo {pages} {1535} (\bibinfo {year}
  {2000})}\BibitemShut {NoStop}%
\bibitem [{\citenamefont {Honecker}\ \emph {et~al.}(2004)\citenamefont
  {Honecker}, \citenamefont {Schulenburg},\ and\ \citenamefont
  {Richter}}]{Richter}%
  \BibitemOpen
  \bibfield  {author} {\bibinfo {author} {\bibfnamefont {A.}~\bibnamefont
  {Honecker}}, \bibinfo {author} {\bibfnamefont {J.}~\bibnamefont
  {Schulenburg}}, \ and\ \bibinfo {author} {\bibfnamefont {J.}~\bibnamefont
  {Richter}},\ }\href {http://stacks.iop.org/0953-8984/16/i=11/a=025}
  {\bibfield  {journal} {\bibinfo  {journal} {J. Phys. Cond. Mat.}\ }\textbf
  {\bibinfo {volume} {16}},\ \bibinfo {pages} {S749} (\bibinfo {year}
  {2004})}\BibitemShut {NoStop}%
\bibitem [{\citenamefont {Lieb}\ \emph {et~al.}(1961)\citenamefont {Lieb},
  \citenamefont {Schultz},\ and\ \citenamefont {Mattis}}]{LiebSchultzMattis}%
  \BibitemOpen
  \bibfield  {author} {\bibinfo {author} {\bibfnamefont {E.}~\bibnamefont
  {Lieb}}, \bibinfo {author} {\bibfnamefont {T.}~\bibnamefont {Schultz}}, \
  and\ \bibinfo {author} {\bibfnamefont {D.}~\bibnamefont {Mattis}},\ }\href
  {\doibase 10.1016/0003-4916(61)90115-4} {\bibfield  {journal} {\bibinfo
  {journal} {Ann. Phys. (N.Y.)}\ }\textbf {\bibinfo {volume} {16}},\ \bibinfo
  {pages} {407 } (\bibinfo {year} {1961})}\BibitemShut {NoStop}%
\bibitem [{\citenamefont {Yan}\ \emph {et~al.}(2011)\citenamefont {Yan},
  \citenamefont {Huse},\ and\ \citenamefont {White}}]{White}%
  \BibitemOpen
  \bibfield  {author} {\bibinfo {author} {\bibfnamefont {S.}~\bibnamefont
  {Yan}}, \bibinfo {author} {\bibfnamefont {D.~A.}\ \bibnamefont {Huse}}, \
  and\ \bibinfo {author} {\bibfnamefont {S.~R.}\ \bibnamefont {White}},\ }\href
  {\doibase 10.1126/science.1201080} {\bibfield  {journal} {\bibinfo  {journal}
  {Science}\ }\textbf {\bibinfo {volume} {332}},\ \bibinfo {pages} {1173}
  (\bibinfo {year} {2011})}\BibitemShut {NoStop}%
\bibitem [{\citenamefont {Depenbrock}\ \emph {et~al.}(2012)\citenamefont
  {Depenbrock}, \citenamefont {McCulloch},\ and\ \citenamefont
  {Schollw\"ock}}]{Depenbrock}%
  \BibitemOpen
  \bibfield  {author} {\bibinfo {author} {\bibfnamefont {S.}~\bibnamefont
  {Depenbrock}}, \bibinfo {author} {\bibfnamefont {I.~P.}\ \bibnamefont
  {McCulloch}}, \ and\ \bibinfo {author} {\bibfnamefont {U.}~\bibnamefont
  {Schollw\"ock}},\ }\href {\doibase 10.1103/PhysRevLett.109.067201} {\bibfield
   {journal} {\bibinfo  {journal} {Phys. Rev. Lett.}\ }\textbf {\bibinfo
  {volume} {109}},\ \bibinfo {pages} {067201} (\bibinfo {year}
  {2012})}\BibitemShut {NoStop}%
\bibitem [{\citenamefont {Jiang}\ \emph
  {et~al.}(2012{\natexlab{a}})\citenamefont {Jiang}, \citenamefont {Wang},\
  and\ \citenamefont {Balents}}]{Balents}%
  \BibitemOpen
  \bibfield  {author} {\bibinfo {author} {\bibfnamefont {H.-C.}\ \bibnamefont
  {Jiang}}, \bibinfo {author} {\bibfnamefont {Z.}~\bibnamefont {Wang}}, \ and\
  \bibinfo {author} {\bibfnamefont {L.}~\bibnamefont {Balents}},\ }\href
  {\doibase 10.1038/nphys2465} {\bibfield  {journal} {\bibinfo  {journal} {Nat.
  Phys.}\ }\textbf {\bibinfo {volume} {8}},\ \bibinfo {pages} {902} (\bibinfo
  {year} {2012}{\natexlab{a}})}\BibitemShut {NoStop}%
\bibitem [{\citenamefont {Fu}\ \emph {et~al.}(2015)\citenamefont {Fu},
  \citenamefont {Imai}, \citenamefont {Han},\ and\ \citenamefont {Lee}}]{Imai}%
  \BibitemOpen
  \bibfield  {author} {\bibinfo {author} {\bibfnamefont {M.}~\bibnamefont
  {Fu}}, \bibinfo {author} {\bibfnamefont {T.}~\bibnamefont {Imai}}, \bibinfo
  {author} {\bibfnamefont {T.-H.}\ \bibnamefont {Han}}, \ and\ \bibinfo
  {author} {\bibfnamefont {Y.~S.}\ \bibnamefont {Lee}},\ }\href {\doibase
  10.1126/science.aab2120} {\bibfield  {journal} {\bibinfo  {journal}
  {Science}\ }\textbf {\bibinfo {volume} {350}},\ \bibinfo {pages} {655}
  (\bibinfo {year} {2015})}\BibitemShut {NoStop}%
\bibitem [{\citenamefont {Jiang}\ \emph
  {et~al.}(2012{\natexlab{b}})\citenamefont {Jiang}, \citenamefont {Yao},\ and\
  \citenamefont {Balents}}]{Hong-Chen}%
  \BibitemOpen
  \bibfield  {author} {\bibinfo {author} {\bibfnamefont {H.-C.}\ \bibnamefont
  {Jiang}}, \bibinfo {author} {\bibfnamefont {H.}~\bibnamefont {Yao}}, \ and\
  \bibinfo {author} {\bibfnamefont {L.}~\bibnamefont {Balents}},\ }\href
  {\doibase 10.1103/PhysRevB.86.024424} {\bibfield  {journal} {\bibinfo
  {journal} {Phys. Rev. B}\ }\textbf {\bibinfo {volume} {86}},\ \bibinfo
  {pages} {024424} (\bibinfo {year} {2012}{\natexlab{b}})}\BibitemShut
  {NoStop}%
\bibitem [{\citenamefont {Affleck}\ and\ \citenamefont
  {Lieb}(1986)}]{Affleck1986}%
  \BibitemOpen
  \bibfield  {author} {\bibinfo {author} {\bibfnamefont {I.}~\bibnamefont
  {Affleck}}\ and\ \bibinfo {author} {\bibfnamefont {E.~H.}\ \bibnamefont
  {Lieb}},\ }\href {\doibase 10.1007/BF00400304} {\bibfield  {journal}
  {\bibinfo  {journal} {Lett. Math. Phys.}\ }\textbf {\bibinfo {volume} {12}},\
  \bibinfo {pages} {57} (\bibinfo {year} {1986})}\BibitemShut {NoStop}%
\bibitem [{\citenamefont {Yamanaka}\ \emph {et~al.}(1997)\citenamefont
  {Yamanaka}, \citenamefont {Oshikawa},\ and\ \citenamefont
  {Affleck}}]{Yamanaka}%
  \BibitemOpen
  \bibfield  {author} {\bibinfo {author} {\bibfnamefont {M.}~\bibnamefont
  {Yamanaka}}, \bibinfo {author} {\bibfnamefont {M.}~\bibnamefont {Oshikawa}},
  \ and\ \bibinfo {author} {\bibfnamefont {I.}~\bibnamefont {Affleck}},\ }\href
  {\doibase 10.1103/PhysRevLett.79.1110} {\bibfield  {journal} {\bibinfo
  {journal} {Phys. Rev. Lett.}\ }\textbf {\bibinfo {volume} {79}},\ \bibinfo
  {pages} {1110} (\bibinfo {year} {1997})}\BibitemShut {NoStop}%
\bibitem [{\citenamefont {Hastings}(2004{\natexlab{a}})}]{HastingsLSM}%
  \BibitemOpen
  \bibfield  {author} {\bibinfo {author} {\bibfnamefont {M.~B.}\ \bibnamefont
  {Hastings}},\ }\href {\doibase 10.1103/PhysRevB.69.104431} {\bibfield
  {journal} {\bibinfo  {journal} {Phys. Rev. B}\ }\textbf {\bibinfo {volume}
  {69}},\ \bibinfo {pages} {104431} (\bibinfo {year}
  {2004}{\natexlab{a}})}\BibitemShut {NoStop}%
\bibitem [{\citenamefont {Parameswaran}\ \emph {et~al.}(2013)\citenamefont
  {Parameswaran}, \citenamefont {Turner}, \citenamefont {Arovas},\ and\
  \citenamefont {Vishwanath}}]{Sid}%
  \BibitemOpen
  \bibfield  {author} {\bibinfo {author} {\bibfnamefont {S.~A.}\ \bibnamefont
  {Parameswaran}}, \bibinfo {author} {\bibfnamefont {A.~M.}\ \bibnamefont
  {Turner}}, \bibinfo {author} {\bibfnamefont {D.~P.}\ \bibnamefont {Arovas}},
  \ and\ \bibinfo {author} {\bibfnamefont {A.}~\bibnamefont {Vishwanath}},\
  }\href {\doibase 10.1038/nphys2600} {\bibfield  {journal} {\bibinfo
  {journal} {Nat. Phys.}\ }\textbf {\bibinfo {volume} {9}},\ \bibinfo {pages}
  {299} (\bibinfo {year} {2013})}\BibitemShut {NoStop}%
\bibitem [{\citenamefont {Watanabe}\ \emph {et~al.}(2015)\citenamefont
  {Watanabe}, \citenamefont {Po}, \citenamefont {Vishwanath},\ and\
  \citenamefont {Zaletel}}]{PNAS}%
  \BibitemOpen
  \bibfield  {author} {\bibinfo {author} {\bibfnamefont {H.}~\bibnamefont
  {Watanabe}}, \bibinfo {author} {\bibfnamefont {H.~C.}\ \bibnamefont {Po}},
  \bibinfo {author} {\bibfnamefont {A.}~\bibnamefont {Vishwanath}}, \ and\
  \bibinfo {author} {\bibfnamefont {M.~P.}\ \bibnamefont {Zaletel}},\ }\href
  {http://www.pnas.org/content/112/47/14551.abstract} {\bibfield  {journal}
  {\bibinfo  {journal} {Proc. Natl. Acad. Sci. U.S.A.}\ }\textbf {\bibinfo
  {volume} {112}},\ \bibinfo {pages} {14551} (\bibinfo {year}
  {2015})}\BibitemShut {NoStop}%
\bibitem [{\citenamefont {Jaksch}\ \emph {et~al.}(1998)\citenamefont {Jaksch},
  \citenamefont {Bruder}, \citenamefont {Cirac}, \citenamefont {Gardiner},\
  and\ \citenamefont {Zoller}}]{Zoller}%
  \BibitemOpen
  \bibfield  {author} {\bibinfo {author} {\bibfnamefont {D.}~\bibnamefont
  {Jaksch}}, \bibinfo {author} {\bibfnamefont {C.}~\bibnamefont {Bruder}},
  \bibinfo {author} {\bibfnamefont {J.~I.}\ \bibnamefont {Cirac}}, \bibinfo
  {author} {\bibfnamefont {C.~W.}\ \bibnamefont {Gardiner}}, \ and\ \bibinfo
  {author} {\bibfnamefont {P.}~\bibnamefont {Zoller}},\ }\href {\doibase
  10.1103/PhysRevLett.81.3108} {\bibfield  {journal} {\bibinfo  {journal}
  {Phys. Rev. Lett.}\ }\textbf {\bibinfo {volume} {81}},\ \bibinfo {pages}
  {3108} (\bibinfo {year} {1998})}\BibitemShut {NoStop}%
\bibitem [{\citenamefont {Batrouni}\ \emph {et~al.}(2002)\citenamefont
  {Batrouni}, \citenamefont {Rousseau}, \citenamefont {Scalettar},
  \citenamefont {Rigol}, \citenamefont {Muramatsu}, \citenamefont {Denteneer},\
  and\ \citenamefont {Troyer}}]{Troyer}%
  \BibitemOpen
  \bibfield  {author} {\bibinfo {author} {\bibfnamefont {G.~G.}\ \bibnamefont
  {Batrouni}}, \bibinfo {author} {\bibfnamefont {V.}~\bibnamefont {Rousseau}},
  \bibinfo {author} {\bibfnamefont {R.~T.}\ \bibnamefont {Scalettar}}, \bibinfo
  {author} {\bibfnamefont {M.}~\bibnamefont {Rigol}}, \bibinfo {author}
  {\bibfnamefont {A.}~\bibnamefont {Muramatsu}}, \bibinfo {author}
  {\bibfnamefont {P.~J.~H.}\ \bibnamefont {Denteneer}}, \ and\ \bibinfo
  {author} {\bibfnamefont {M.}~\bibnamefont {Troyer}},\ }\href {\doibase
  10.1103/PhysRevLett.89.117203} {\bibfield  {journal} {\bibinfo  {journal}
  {Phys. Rev. Lett.}\ }\textbf {\bibinfo {volume} {89}},\ \bibinfo {pages}
  {117203} (\bibinfo {year} {2002})}\BibitemShut {NoStop}%
\bibitem [{\citenamefont {Bak}(1982)}]{cireview}%
  \BibitemOpen
  \bibfield  {author} {\bibinfo {author} {\bibfnamefont {P.}~\bibnamefont
  {Bak}},\ }\href {http://stacks.iop.org/0034-4885/45/i=6/a=001} {\bibfield
  {journal} {\bibinfo  {journal} {Rep. Prog. Phys.}\ }\textbf {\bibinfo
  {volume} {45}},\ \bibinfo {pages} {587} (\bibinfo {year} {1982})}\BibitemShut
  {NoStop}%
\bibitem [{\citenamefont {Altland}\ and\ \citenamefont
  {Simons}(2010)}]{Altland}%
  \BibitemOpen
  \bibfield  {author} {\bibinfo {author} {\bibfnamefont {A.}~\bibnamefont
  {Altland}}\ and\ \bibinfo {author} {\bibfnamefont {B.}~\bibnamefont
  {Simons}},\ }\href@noop {} {\emph {\bibinfo {title} {{Condensed Matter Field
  Theory}}}},\ \bibinfo {edition} {2nd}\ ed.\ (\bibinfo  {publisher} {Cambridge
  University Press, Cambridge, England},\ \bibinfo {year} {2010})\BibitemShut
  {NoStop}%
\bibitem [{\citenamefont {Mahan}(2000)}]{Mahan}%
  \BibitemOpen
  \bibfield  {author} {\bibinfo {author} {\bibfnamefont {G.~D.}\ \bibnamefont
  {Mahan}},\ }\href@noop {} {\emph {\bibinfo {title} {{Many-Particle
  Physics}}}},\ \bibinfo {edition} {3rd}\ ed.\ (\bibinfo  {publisher} {Plenum
  Publishers, New York},\ \bibinfo {year} {2000})\BibitemShut {NoStop}%
\bibitem [{\citenamefont {Horsch}\ and\ \citenamefont {von~der
  Linden}(1988)}]{HorschvonderLinden}%
  \BibitemOpen
  \bibfield  {author} {\bibinfo {author} {\bibfnamefont {P.}~\bibnamefont
  {Horsch}}\ and\ \bibinfo {author} {\bibfnamefont {W.}~\bibnamefont {von~der
  Linden}},\ }\href {\doibase 10.1007/BF01312134} {\bibfield  {journal}
  {\bibinfo  {journal} {Z. Phys. B}\ }\textbf {\bibinfo {volume} {72}},\
  \bibinfo {pages} {181} (\bibinfo {year} {1988})}\BibitemShut {NoStop}%
\bibitem [{\citenamefont {Koma}\ and\ \citenamefont
  {Tasaki}(1994)}]{KomaTasaki}%
  \BibitemOpen
  \bibfield  {author} {\bibinfo {author} {\bibfnamefont {T.}~\bibnamefont
  {Koma}}\ and\ \bibinfo {author} {\bibfnamefont {H.}~\bibnamefont {Tasaki}},\
  }\href {\doibase 10.1007/BF02188685} {\bibfield  {journal} {\bibinfo
  {journal} {J. Stat. Phys.}\ }\textbf {\bibinfo {volume} {76}},\ \bibinfo
  {pages} {745} (\bibinfo {year} {1994})}\BibitemShut {NoStop}%
\bibitem [{\citenamefont {Watanabe}\ and\ \citenamefont
  {Oshikawa}(2015)}]{WatanabeOshikawa}%
  \BibitemOpen
  \bibfield  {author} {\bibinfo {author} {\bibfnamefont {H.}~\bibnamefont
  {Watanabe}}\ and\ \bibinfo {author} {\bibfnamefont {M.}~\bibnamefont
  {Oshikawa}},\ }\href {\doibase 10.1103/PhysRevLett.114.251603} {\bibfield
  {journal} {\bibinfo  {journal} {Phys. Rev. Lett.}\ }\textbf {\bibinfo
  {volume} {114}},\ \bibinfo {pages} {251603} (\bibinfo {year}
  {2015})}\BibitemShut {NoStop}%
\bibitem [{\citenamefont {Else}\ \emph {et~al.}(2016)\citenamefont {Else},
  \citenamefont {Bauer},\ and\ \citenamefont {Nayak}}]{Nayak}%
  \BibitemOpen
  \bibfield  {author} {\bibinfo {author} {\bibfnamefont {D.~V.}\ \bibnamefont
  {Else}}, \bibinfo {author} {\bibfnamefont {B.}~\bibnamefont {Bauer}}, \ and\
  \bibinfo {author} {\bibfnamefont {C.}~\bibnamefont {Nayak}},\ }\href
  {\doibase 10.1103/PhysRevLett.117.090402} {\bibfield  {journal} {\bibinfo
  {journal} {Phys. Rev. Lett.}\ }\textbf {\bibinfo {volume} {117}},\ \bibinfo
  {pages} {090402} (\bibinfo {year} {2016})}\BibitemShut {NoStop}%
\bibitem [{\citenamefont {Hastings}(2004{\natexlab{b}})}]{Hastings}%
  \BibitemOpen
  \bibfield  {author} {\bibinfo {author} {\bibfnamefont {M.~B.}\ \bibnamefont
  {Hastings}},\ }\href {\doibase 10.1103/PhysRevLett.93.140402} {\bibfield
  {journal} {\bibinfo  {journal} {Phys. Rev. Lett.}\ }\textbf {\bibinfo
  {volume} {93}},\ \bibinfo {pages} {140402} (\bibinfo {year}
  {2004}{\natexlab{b}})}\BibitemShut {NoStop}%
\bibitem [{\citenamefont {Hastings}\ and\ \citenamefont
  {Koma}(2006)}]{HastingsKoma}%
  \BibitemOpen
  \bibfield  {author} {\bibinfo {author} {\bibfnamefont {M.~B.}\ \bibnamefont
  {Hastings}}\ and\ \bibinfo {author} {\bibfnamefont {T.}~\bibnamefont
  {Koma}},\ }\href {\doibase 10.1007/s00220-006-0030-4} {\bibfield  {journal}
  {\bibinfo  {journal} {Commun. Math. Phys.}\ }\textbf {\bibinfo {volume}
  {265}},\ \bibinfo {pages} {781} (\bibinfo {year} {2006})}\BibitemShut
  {NoStop}%
\bibitem [{\citenamefont {Frahm}\ and\ \citenamefont {Sobiella}(1999)}]{Frahm}%
  \BibitemOpen
  \bibfield  {author} {\bibinfo {author} {\bibfnamefont {H.}~\bibnamefont
  {Frahm}}\ and\ \bibinfo {author} {\bibfnamefont {C.}~\bibnamefont
  {Sobiella}},\ }\href {\doibase 10.1103/PhysRevLett.83.5579} {\bibfield
  {journal} {\bibinfo  {journal} {Phys. Rev. Lett.}\ }\textbf {\bibinfo
  {volume} {83}},\ \bibinfo {pages} {5579} (\bibinfo {year}
  {1999})}\BibitemShut {NoStop}%
\bibitem [{\citenamefont {Cabra}\ \emph {et~al.}(2000)\citenamefont {Cabra},
  \citenamefont {Martino}, \citenamefont {Honecker}, \citenamefont {Pujol},\
  and\ \citenamefont {Simon}}]{Cabra2000}%
  \BibitemOpen
  \bibfield  {author} {\bibinfo {author} {\bibfnamefont {D.}~\bibnamefont
  {Cabra}}, \bibinfo {author} {\bibfnamefont {A.~D.}\ \bibnamefont {Martino}},
  \bibinfo {author} {\bibfnamefont {A.}~\bibnamefont {Honecker}}, \bibinfo
  {author} {\bibfnamefont {P.}~\bibnamefont {Pujol}}, \ and\ \bibinfo {author}
  {\bibfnamefont {P.}~\bibnamefont {Simon}},\ }\href {\doibase
  http://dx.doi.org/10.1016/S0375-9601(00)00210-3} {\bibfield  {journal}
  {\bibinfo  {journal} {Phys. Lett. A}\ }\textbf {\bibinfo {volume} {268}},\
  \bibinfo {pages} {418 } (\bibinfo {year} {2000})}\BibitemShut {NoStop}%
\bibitem [{\citenamefont {Cabra}\ \emph {et~al.}(2002)\citenamefont {Cabra},
  \citenamefont {Martino}, \citenamefont {Pujol},\ and\ \citenamefont
  {Simon}}]{Cabra2002}%
  \BibitemOpen
  \bibfield  {author} {\bibinfo {author} {\bibfnamefont {D.}~\bibnamefont
  {Cabra}}, \bibinfo {author} {\bibfnamefont {A.~D.}\ \bibnamefont {Martino}},
  \bibinfo {author} {\bibfnamefont {P.}~\bibnamefont {Pujol}}, \ and\ \bibinfo
  {author} {\bibfnamefont {P.}~\bibnamefont {Simon}},\ }\href {\doibase
  https://doi.org/10.1209/epl/i2002-00475-5} {\bibfield  {journal} {\bibinfo
  {journal} {Europhys. Lett.}\ }\textbf {\bibinfo {volume} {57}},\ \bibinfo
  {pages} {402} (\bibinfo {year} {2002})}\BibitemShut {NoStop}%
\bibitem [{\citenamefont {Roux}\ \emph {et~al.}(2006)\citenamefont {Roux},
  \citenamefont {White}, \citenamefont {Capponi},\ and\ \citenamefont
  {Poilblanc}}]{Roux}%
  \BibitemOpen
  \bibfield  {author} {\bibinfo {author} {\bibfnamefont {G.}~\bibnamefont
  {Roux}}, \bibinfo {author} {\bibfnamefont {S.~R.}\ \bibnamefont {White}},
  \bibinfo {author} {\bibfnamefont {S.}~\bibnamefont {Capponi}}, \ and\
  \bibinfo {author} {\bibfnamefont {D.}~\bibnamefont {Poilblanc}},\ }\href
  {\doibase 10.1103/PhysRevLett.97.087207} {\bibfield  {journal} {\bibinfo
  {journal} {Phys. Rev. Lett.}\ }\textbf {\bibinfo {volume} {97}},\ \bibinfo
  {pages} {087207} (\bibinfo {year} {2006})}\BibitemShut {NoStop}%
\bibitem [{\citenamefont {Lamas}\ \emph {et~al.}(2011)\citenamefont {Lamas},
  \citenamefont {Capponi},\ and\ \citenamefont {Pujol}}]{Lamas}%
  \BibitemOpen
  \bibfield  {author} {\bibinfo {author} {\bibfnamefont {C.~A.}\ \bibnamefont
  {Lamas}}, \bibinfo {author} {\bibfnamefont {S.}~\bibnamefont {Capponi}}, \
  and\ \bibinfo {author} {\bibfnamefont {P.}~\bibnamefont {Pujol}},\ }\href
  {\doibase 10.1103/PhysRevB.84.115125} {\bibfield  {journal} {\bibinfo
  {journal} {Phys. Rev. B}\ }\textbf {\bibinfo {volume} {84}},\ \bibinfo
  {pages} {115125} (\bibinfo {year} {2011})}\BibitemShut {NoStop}%
\bibitem [{\citenamefont {Nishimori}\ and\ \citenamefont
  {Miyashita}(1986)}]{Miyashita}%
  \BibitemOpen
  \bibfield  {author} {\bibinfo {author} {\bibfnamefont {H.}~\bibnamefont
  {Nishimori}}\ and\ \bibinfo {author} {\bibfnamefont {S.}~\bibnamefont
  {Miyashita}},\ }\href {\doibase 10.1143/JPSJ.55.4448} {\bibfield  {journal}
  {\bibinfo  {journal} {J. Phys. Soc. Jpn}\ }\textbf {\bibinfo {volume} {55}},\
  \bibinfo {pages} {4448} (\bibinfo {year} {1986})}\BibitemShut {NoStop}%
\end{thebibliography}%
\clearpage

\end{document}